\documentclass[showpacs,pra,twocolumn,amssymb]{revtex4}
\usepackage{graphicx}

\begin{document}

\title{Critical Zeeman Splitting of a Unitary Fermi Superfluid}
\author{Lianyi He$^{1,2}$ and Pengfei Zhuang$^{2}$}

\affiliation{1 Frankfurt Institute for Advanced Studies and Institute
for Theoretical Physics, Johann Wolfgang Goethe University, D-60438 Frankfurt am
Main, Germany\\
2 Department of Physics, Tsinghua University, Beijing
100084, China}

\begin{abstract}
We determine the critical Zeeman energy splitting of a homogeneous Fermi superfluid at unitary
in terms of the Fermi energy $\epsilon_{\text F}$ according to recent experimental results in LKB-Lhomond.
Based on the universal equations of state for the superfluid and normal phases, we show that there exist two
critical fields $H_{c1}$ and $H_{c2}$, between which a
superfluid-normal mixed phase is energetically favored.
Universal formulae for the critical fields and the critical
population imbalance $P_c$ are derived. We have found a universal relation between the critical fields and the critical imbalances:
$H_{c1}=\gamma\xi\epsilon_{\text F}$ and $H_{c2}=(1+\gamma P_c)^{2/3}H_{c1}$ where $\xi$ is the universal constant and $\gamma$ is the critical value of the chemical potential imbalance
in the grand canonical ensemble.
Since $\xi$, $\gamma$ and $P_c$ have been measured in the experiments, we can determine the critical Zeeman fields without the detail information of the
equation of state for the polarized normal phase. Using the experimental data from LKB-Lhomond, we have found $H_{c1}\simeq0.37\epsilon_{\text F}$ and $H_{c2}\simeq0.44\epsilon_{\text F}$.
Our result of the polarization $P$ as a function of the Zeeman field $H/\epsilon_{\text F}$ is in good agreement with the data extracted from the experiments.
We also give an estimation of the critical magnetic field for dilute neutron matter at which the matter gets spin polarized, assuming the properties of the
dilute neutron matter are close to those of the unitary Fermi gas.
\end{abstract}

 \maketitle

%%%%%%%%%%%%%%%%%%%%%%%%%%%%%%%%%%%%%%%%%%%%%%%%%%%%%%%%%%%%%%%%%%%%%%%
\section {Introduction}
\label{s1}
%%%%%%%%%%%%%%%%%%%%%%%%%%%%%%%%%%%%%%%%%%%%%%%%%%%%%%%%%%%%%%%%%%%%%%%

While Bardeen-Cooper-Schiffer (BCS)
superconductivity/superfluidity~\cite{BCS} in Fermi systems was
investigted more than 50 years ago, the main scientific interest
in recent experiments of two-component ultracold Fermi gas is to
create a new type of Fermi superfluid in the BCS--Bose-Einstein condensation
(BEC) crossover~\cite{BCSBEC,BCSBEC02,BCSBEC03,BCSBEC04,BCSBECexp,BCSBECexp02,BCSBECexp03}. At the Feshbach resonance
where the s-wave scattering length $a_s$ diverges, a novel type of
Fermi superfluid has been observed~\cite{BCSBECexp,BCSBECexp02,BCSBECexp03}. In the dilute gas limit, where the finite-range
effect of the interaction can be neglected, the only length scale in the many-body problem is the
inter-particle distance. In this so-called unitary
Fermi gas, any physical quantity can be expressed in terms of its
value in the non-interacting case multiplied by a universal
constant~\cite{ho}. For example, the energy density can be written
as ${\cal E}=\xi {\cal E}_0$, where $\xi$ is a universal constant
and ${\cal E}_0$ the energy density of the non-interacting Fermi gas. A possible
realization of such a unitary Fermi superfluid is the dilute neutron matter~\cite{DNM,DNM02,DNM03} which exists
in neutron star crust, since the neutron-neutron scattering length $a_{nn}\simeq-18.5$fm is much larger
than the effective range of the nuclear force and the inter-particle distance.

In addition to the idealized case where fermion pairing occurs on
a uniform Fermi surface, the effect of pure Zeeman energy splitting
$E_Z=2H$ (Here $H=\mu_{\text B}B$ with the effective ``magnetic field" $B$ and the magnetic moment $\mu_{\text B}$)
between spin-up and -down electrons in the BCS
superconductivity was known many years ago~\cite{CC,CC2,Sarma,FFLO,FFLO2}. At a critical Zeeman
field or the so-called Chandrasekhar-Clogston (CC) limit
$H_c=0.707\Delta_0$ where $\Delta_0$ is the zero temperature gap,
a first order phase transition from the gapped BCS state to the
normal state occurs. Further theoretical studies showed that the
inhomogeneous Fulde-Ferrell-Larkin-Ovchinnikov (FFLO)
state~\cite{FFLO, FFLO2} may survive in a narrow window between $H_c$ and
$H_{\text{FFLO}}=0.754\Delta_0$. However, since the thermodynamic
critical field is much smaller than the CC limit due to strong
orbit effect \cite{CC}, it is hard to observe the CC limit and the
FFLO state in ordinary superconductors.

Recent experiments for strongly interacting ultracold Fermi gases
give an alternative way to study the pure Zeeman effect on Fermi
superfluidity~\cite{imbalanceexp, imbalanceexp2}. The atom numbers of the two
lowest hyperfine states of $^6$Li, denoted by $N_\uparrow$ and
$N_\downarrow$, are adjusted to create a population imbalance,
which simulates effectively the Zeeman field $H$ in a
superconductor. At the unitary point, phase separation phenomenon between the unpolarized
superfluid and the polarized normal gas, predicted by early theoretical
works~\cite{BCR,Cohen} and Quantum Monte Carlo (QMC)
calculations~\cite{Carlson}, has been observed~\cite{imbalanceexp, imbalanceexp2}. However, the evidences for the
FFLO and the breached pairing states~\cite{BP,he1} have not yet been found at unitary. The problem of
imbalanced pairing in strongly interacting Fermi systems is also relevant to nuclear matter~\cite{nuclear} and
quark matter~\cite{quark} which may exist in compact stars. Since the neutrons carry a tiny magnetic moment
$\mu_n\simeq6.03\times10^{-18}$MeV$\cdot$G$^{-1}$~(in natural units), the superfluid state in dilute neutron matter can be destroyed
when the magnetic field reaches a critical value and the matter becomes spin-polarized~\cite{qmcn}.
Since the effective interaction in dilute neutron matter
is quite close to unitary, the experimental data from cold atoms can help us determine the critical magnetic field
for dilute neutron matter.

Due to the universality at infinite scattering length,
the CC limit of a unitary Fermi superfluid should be a universal constant. For cold atom systems, the chemical potentials for the
$\uparrow$ and $\downarrow$ atoms are different. They can be denoted as $\mu_{\uparrow}$ and $\mu_{\downarrow}$.
We define the averaged chemical potential $\mu=(\mu_\uparrow+\mu_\downarrow)/2$
and the Zeeman field $H=(\mu_\uparrow-\mu_\downarrow)/2$. For condensed matter systems, the Zeeman field $H$ is induced by some external field.
There exist two types of CC limit which can be directly measured in experiments or QMC calculations
corresponding to different ensembles :
\\
(1)The critical value $\gamma$ for the chemical potential imbalance
\begin{equation}
\frac{H}{\mu}=\frac{\mu_{\uparrow}-\mu_{\downarrow}}{\mu_{\uparrow}+\mu_{\downarrow}}
\end{equation}
at which a first order phase transition between the superfluid and the normal phases takes place.
\\
(2)The critical value $P_c$ for the population imbalance
\begin{equation}
P=\frac{N_{\uparrow}-N_{\downarrow}}{N_{\uparrow}+N_{\downarrow}}\label{defP}
\end{equation}
above which the superfluid-normal mixed phase disappears. These two critical values have been successfully
determined in experiments and QMC calculations in recent years~\cite{MC,exp,exp2,exp3}.

To make a closer connection with condensed matter
systems and cold dilute neutron matter in neutron star crust, we should convert the results from cold atoms
to usual variables of condensed matter systems,
i.e., the critical Zeeman energy splitting $H_c$ in terms of the Fermi energy $\epsilon_{\text F}=(3\pi^2 n)^{2/3}/(2M)$~($M$ is
the fermion mass)
for a homogeneous Fermi gas with fixed total density $n$. Obviously, the ratio $H_c/\epsilon_{\text F}$
should be a universal constant at unitary. However, from the simple mean-field approach in
Ref.~\cite{sheehy,sheehy2,he2} it was found that there should exist two critical
fields $H_{c1}$ and $H_{c2}$ in the BCS-BEC crossover and the superfluid-normal mixed phase
is the energetically favored ground state in the region
of $H_{c1}<H<H_{c2}$. For $H<H_{c1}$, the ground state is an unpolarized superfluid with balanced spin populations.
The matter becomes a spin-polarized normal Fermi liquid for $H>H_{c2}$.  The existence of the two critical fields is somewhat like the type-II superconductors,
but mechanism is different. Since the
particle numbers $N_\uparrow$ and $N_\downarrow$ are used as tunable
parameters in QMC calculations and experiments, only the first order
phase transition point $\gamma=(H/\mu)_c$ and the critical population imbalance
$P_c$ have been directly determined. The two critical fields $H_{c1}$ and $H_{c2}$ in terms of the Fermi energy $\epsilon_{\text F}$
for homogeneous Fermi gases have not yet been determined by QMC
calculations and experiments.

In this paper, we will determine the two critical fields, $H_{c1}$ and
$H_{c2}$, in terms of the Fermi energy $\epsilon_{\text F}$ for a
homogeneous Fermi gas at infinite scattering length, based on the
universal properties of the thermodynamics. In our study, we use the general equations of state for the superfluid and
the normal phases~\cite{forbes,chevy}. We neither assume that
the normal phase is fully polarized~\cite{Carlson,son} nor adopt the non-interacting equation of state for the
partially polarized normal phase in the naive mean-field approach~\cite{sheehy2}. Our results are therefore reliable
once the QMC or experimental data are used as input.

%%%%%%%%%%%%%%%%%%%%%%%%%%%%%%%%%%%%%%%%%%%%%%%%%%%%%%%%%%%%%%%%%%%%%%%
\section {Formalism}
\label{s2}
%%%%%%%%%%%%%%%%%%%%%%%%%%%%%%%%%%%%%%%%%%%%%%%%%%%%%%%%%%%%%%%%%%%%%%%

%%%%%%%%%%%%%%%%%%%%%%%%%%%%%%%%%%%%%%%%%%%%%%%%%%%%%%%%%%%%%%%%%%%%%%%
\subsection {Equations of State and Phase Transition}
\label{s2-1}
%%%%%%%%%%%%%%%%%%%%%%%%%%%%%%%%%%%%%%%%%%%%%%%%%%%%%%%%%%%%%%%%%%%%%%%
At unitary, we can construct the exact equations of state (EOS) in
the grand canonical ensemble from the universality hypothesis. The
pressure of the polarized normal phase (N) as a function of the
averaged chemical potential $\mu$ and the Zeeman field $H$ takes the form~\cite{ho}
\begin{equation}
{\cal P}_{\text N}(\mu,H)={\cal P}_0(\mu){\cal
G}\left(\frac{H}{\mu}\right),
\end{equation}
where ${\cal P}_0(\mu)=\frac{2}{5}c\mu^{5/2}$ is the pressure
of the non-interacting Fermi gas with $c=(2M)^{3/2}/(3\pi^2)$, and ${\cal G}(x)$ is a universal scaling
function representing the strong coupling effect and can be determined from the experiments and
QMC calculations. For the balanced case, we define ${\cal G}(0)=\xi_{\text N}^{-3/2}$ and the universal constant
$\xi_{\text N}$ for the normal phase should be larger than that for the superfluid phase, $\xi_{\text N}>\xi$, representing the fact that
the superfluid state is the ground state at $H=0$. Experiments and QMC calculations have determined the value
$\xi_{\text N}\simeq0.51-0.56$~\cite{MC,exp2}. The total number density and
the spin density imbalance, i.e., $n_{\text N}(\mu,H)=\partial{\cal
P}_{\text N}/\partial\mu$ and $m_{\text N}(\mu,H)=\partial{\cal
P}_{\text N}/\partial H$, are given by
\begin{eqnarray}
n_{\text N}(\mu,H)&=&n_0(\mu)\left[{\cal G}\left(\frac{H}{\mu}\right)-\frac{2}{5}\frac{H}{\mu}{\cal G}^\prime\left(\frac{H}{\mu}\right)\right],\nonumber\\
m_{\text N}(\mu,H)&=&n_0(\mu)\frac{2}{5}{\cal G}^\prime\left(\frac{H}{\mu}\right),\label{numn}
\end{eqnarray}
where $n_0(\mu)=c\mu^{3/2}$ is the density of a non-interacting Fermi gas.

There exists a critical value $\delta_0$ of the ratio $H/\mu$, above which the normal
Fermi gas is in the fully polarized state (N$_{\text{FP}}$) with a single spin component $\uparrow$, i.e.,
$m_{\text N}=n_{\text N}$ or $P=1$. In this case the Fermi gas should be non-interacting and the
scaling function is given by ${\cal G}(x)=\frac{1}{2}(1+x)^{5/2}$. For $H/\mu<\delta_0$,
the normal phase is partially polarized (N$_{\text{PP}}$) with $m_{\text N}<n_{\text N}$.
While the mean-field theory predicts $\delta_0=1$~\cite{sheehy}, it is $3.78$~\cite{MC} from
recent QMC calculations. Theoretically, the numerical value of $\delta_0$ can be determined by studying the impurity
problem of a single $\downarrow$ atom immersed in a Fermi sea of $\uparrow$ atoms~\cite{fulmu}.

%%%%%%%%%%%%%%%%%%%%%%%%%%%%%%%%%%%%%%%%%%%%%%%%%%%%%%%%%%%%%%%%%%%%%%%
\begin{figure}[!htb]
\begin{center}
\includegraphics[width=8.5cm]{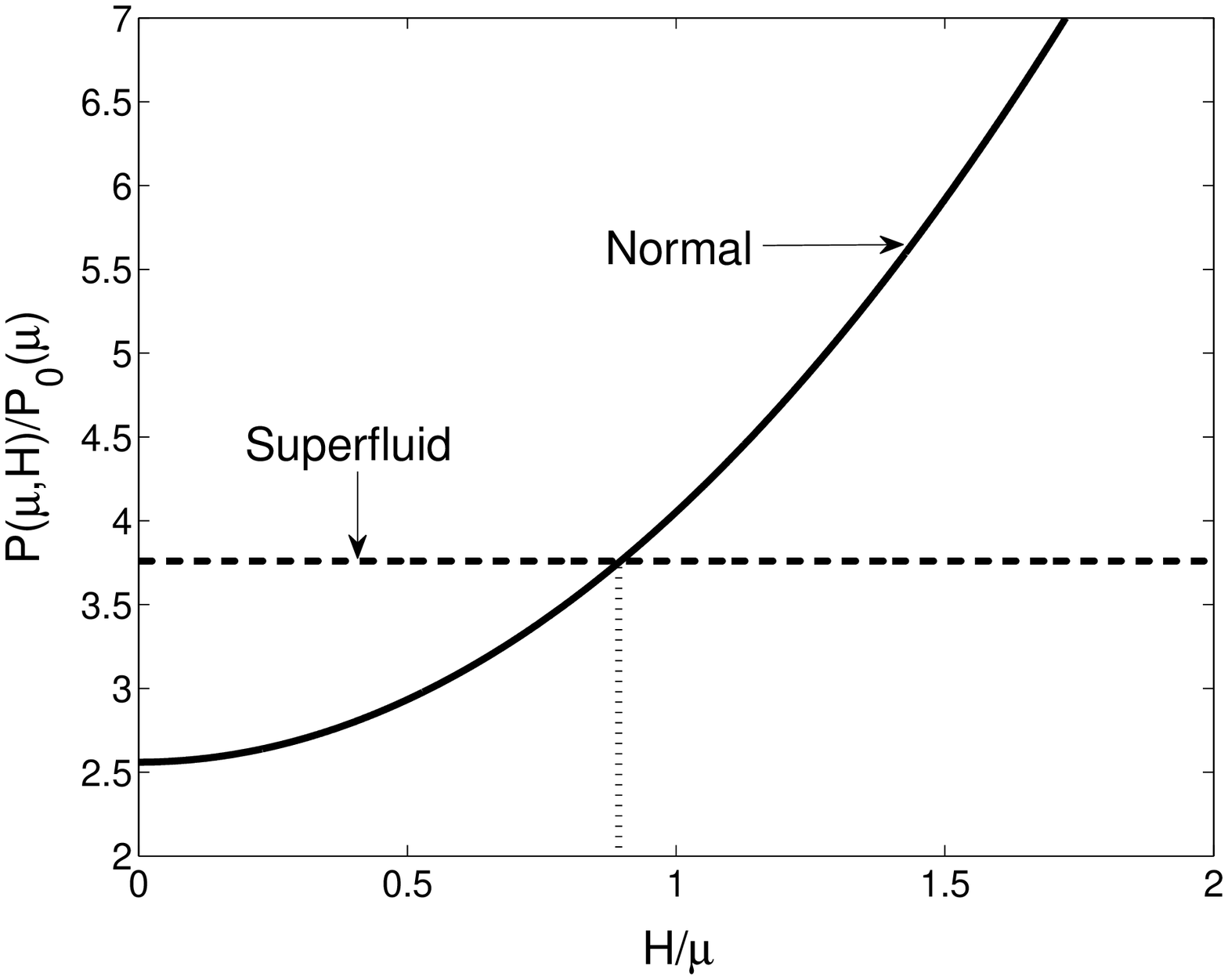}
\includegraphics[width=8.5cm]{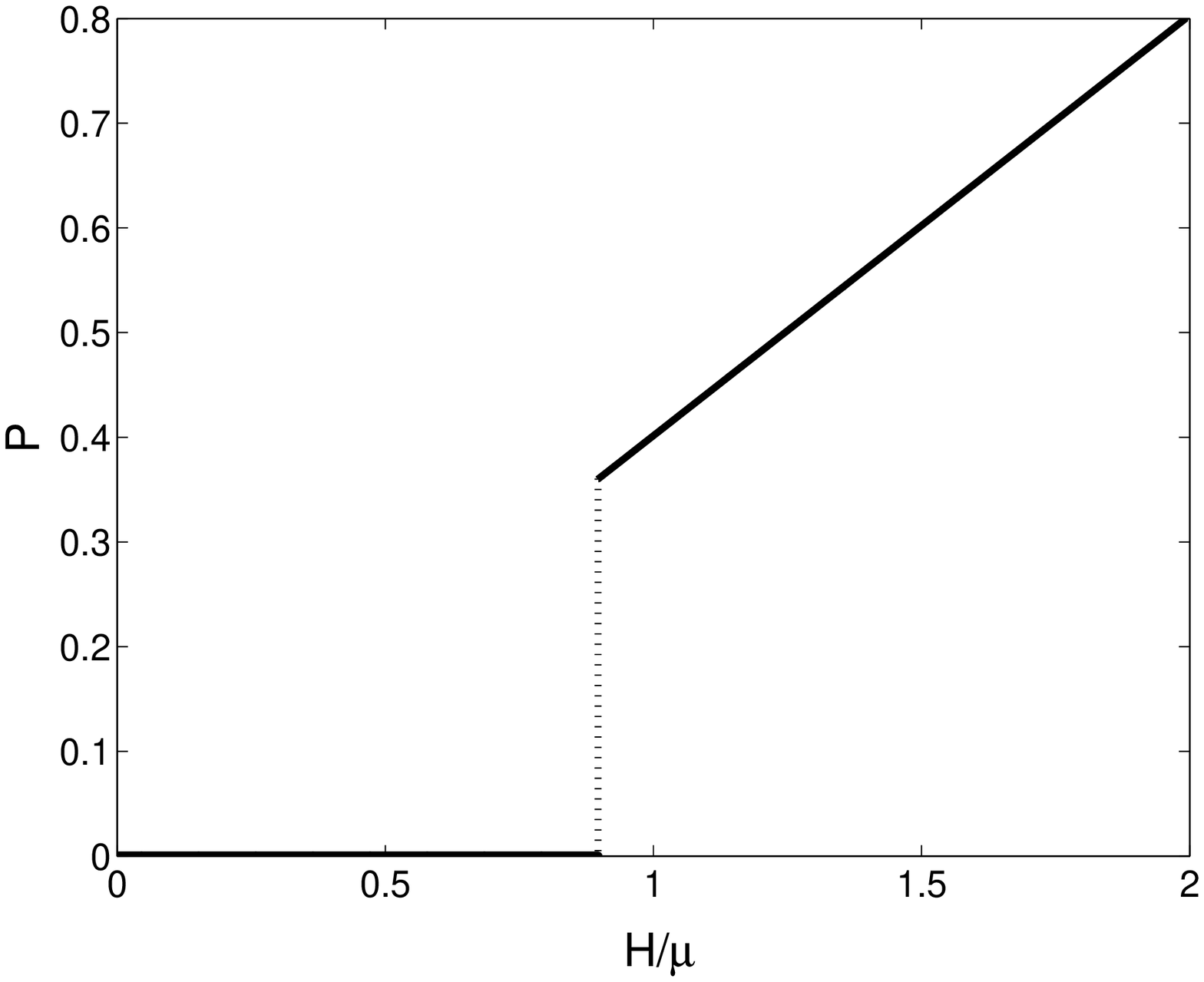}
\caption{(a) Schematic plot of the equations of state for the normal and superfluid states. The vertical dotted line
shows the first order phase transition point $H/\mu=\gamma$. (b) The spin polarization $P$ as a function of $H/\mu$.
As we expect from the first order phase transition, there is a jump from $P=0$ to $P=P_c$ at the first order phase transition point
$H/\mu=\gamma$. \label{fig1}}
\end{center}
\end{figure}
%%%%%%%%%%%%%%%%%%%%%%%%%%%%%%%%%%%%%%%%%%%%%%%%%%%%%%%%%%%%%%%%%%%%%%%%

QMC studies \cite{Carlson,MC} and experiments \cite{exp,exp2,exp3} have shown that
the stable superfluid phase (SF) at unitary should be unpolarized.
Hence the pressure does not depend on $H$ explicitly and takes the
well-known form
\begin{equation}
{\cal P}_{\text {SF}}(\mu,H)={\cal P}_0(\mu)\xi^{-3/2}.
\end{equation}
The total density reads $n_{\text
{SF}}(\mu,H)=n_0(\mu)\xi^{-3/2}$.

Considering the relation $\xi_{\text N}>\xi$ and the fact that ${\cal G}(x)$ is an increasing function of $x$, a first order phase
transition takes place when the pressures of the normal and superfluid phases become equal, i.e., when $H/\mu$ reaches another universal
critical value $\gamma$, which is determined by the following equation
\begin{equation}
\label{gr}
{\cal G}\left(\gamma\right)=\xi^{-3/2}.
\end{equation}
For the case $\gamma>\delta_0$, i.e., the normal phase at the phase transition is N$_{\text{FP}}$, which was assumed in the
early studies \cite{Carlson,son}, we
have $\frac{1}{2}\left(1+\gamma\right)^{5/2}=\xi^{-3/2}$. Combining with
the relation for the excitation gap $\Delta(\mu)=\beta\mu$ at unitary for the balanced case ($H=0$) with a universal constant $\beta$ , we
recover the CC limit derived in~\cite{Carlson}
\begin{equation}
\frac{H_c}{\Delta(\mu)}=\frac{1}{\beta}\left(\frac{2^{2/5}}{\xi^{3/5}}-1\right).\label{carl}
\end{equation}
With the QMC data $\xi=0.42(1)$ and $\beta=1.2(1)$, the authors in \cite{Carlson} obtained $H_c\simeq1.00(5)\Delta$~\cite{Carlson}. However, in this case we found
$\gamma=1.22<\delta_0\simeq3.78$, using the same value of $\xi$ adopted in \cite{Carlson}. This means that the normal phase should be
partially polarized at the phase transition and hence the critical population imbalance $P_c$ should be smaller than unity.
Therefore, the assumption of fully polarized normal state is not correct at unitary.
In Fig. \ref{fig1}, we demonstrate a schematic plot of the equations of state for the superfluid and normal phases, assuming that the
normal phase is partially polarized at the first order phase transition.

%%%%%%%%%%%%%%%%%%%%%%%%%%%%%%%%%%%%%%%%%%%%%%%%%%%%%%%%%%%%%%%%%%%%%%%
\subsection {Critical Zeeman Fields for Homogeneous System}
\label{s2-2}
%%%%%%%%%%%%%%%%%%%%%%%%%%%%%%%%%%%%%%%%%%%%%%%%%%%%%%%%%%%%%%%%%%%%%%%
To make a closer connection with condensed matter
systems and cold dilute neutron matter in neutron star crust, we should consider a system
where the total density $n$ is fixed and the Zeeman splitting $H$
is induced by a tunable external field. For neutron matter, the Zeeman splitting
can be induced by a strong external magnetic field, since the neutrons carry a tiny
magnetic moment.  To determine the critical fields for homogeneous
Fermi superfluids, we turn to the canonical ensemble with fixed total
particle number density $n=(2M\epsilon_{\text F})^{3/2}/(3\pi^2)$.
The Zeeman splitting $H$ is treated as a real external field, and
the conversion between particles in the states $\uparrow$ and
$\downarrow$ is allowed, but the chemical potential $\mu$ is
not a free parameter.

The chemical potential $\mu_{\text N}(H)$ in the polarized normal
phase is solved from the number equation $n_{\text N}(\mu_{\text
N},H)=n$, and the energy density ${\cal E}_{\text N}(H)=\mu_{\text
N}(H)n-{\cal P}_{\text N}(\mu_{\text N}(H),H)$ reads
\begin{equation}
{\cal E}_{\text N}(H)=\frac{5}{3}\left[\frac{\mu_{\text
N}}{\epsilon_{\text F}}-\frac{2}{5}{\cal
G}\left(\frac{H}{\mu_{\text N}}\right)\left(\frac{\mu_{\text
N}}{\epsilon_{\text F}}\right)^{5/2}\right]{\cal E}_0
\end{equation}
with ${\cal E}_0=\frac{3}{5}c\epsilon_{\text F}^{5/2}$. At $H=0$, we have
$\mu_{\text N}(0)=\xi_{\text N}\epsilon_{\text F}$ and ${\cal E}_{\text N}(0)=\xi_{\text N}{\cal E}_0$. For nonzero $H$, the
well-known relation ${\cal E}=3{\cal P}/2$~\cite{ho} is broken, since the interacting energy with the external field $H$ is
included. In the fully polarized normal phase, we have $\mu_{\text
N}(H)=2^{2/3}\epsilon_{\text F}-H$. The
N$_{\text{PP}}$-N$_{\text{FP}}$ transition occurs at
$H_0=2^{2/3}\epsilon_{\text F}\delta_0/(1+\delta_0)$. While the mean-field theory predicts $H_0=2^{-1/3}\epsilon_{\text
F}\simeq0.794\epsilon_{\text F}$, we obtain
$H_0\simeq1.26\epsilon_{\text F}$ from the QMC result $\delta_0\simeq3.78$ \cite{MC}.

Solving the number equation $n_{\text{SF}}(\mu,H)=n$ for the
superfluid phase, the chemical potential and energy density are
given by $\mu_{\text{SF}}(H)=\xi\epsilon_{\text F}$ and ${\cal
E}_{\text{SF}}(H)=\xi {\cal E}_0$, respectively. At $H=0$, there is the BCS
instability ${\cal E}_{\text{SF}}(0)<{\cal E}_{\text{N}}(0)$ which
is numerically supported by the fact $\xi_{\text N}>\xi$. While
${\cal E}_{\text{SF}}(H)$ keeps independent of $H$, ${\cal
E}_{\text{N}}(H)$ should be a monotonously decreasing function. If
there exists no heterogeneous mixed phase, a phase transition
occurs at ${\cal E}_{\text{N}}(H_c) = {\cal E}_{\text{SF}}$. Thus $H_c$ can be determined once
${\cal G}(x)$ is known. Furthermore, if we assume that the
normal state at $H_c$ is fully polarized, i.e., $H_c\geq H_0$, we
obtain $H_c=\frac{3}{5}(2^{2/3}-\xi)\epsilon_{\text F}$.
Using the QMC data $\xi\simeq0.42$, we get $H_c\simeq0.7\epsilon_{\text F}$ which is smaller than $H_0$.
According to the QMC data for the excitation gap $\Delta_0\simeq0.4-0.5\epsilon_{\text F}$ \cite{gap01,gap02}, it is
also in contradiction to the constraint $H_c<\Delta_0$ which ensures that the superfluid
phase is unpolarized. Therefore, the above assumption of phase transition between SF and N$_{\text{FP}}$ without
SF-N$_{\text{FP}}$ mixed phase is not correct.

The above simple analysis without heterogeneous SF-N mixed phase is
not adequate since the first order phase transition should be
associated with the phase separation phenomenon when the total density $n$ is fixed.
Now we take the heterogeneous SF-N mixed phase into account. In fact, due to the strong coupling effect in the
BCS-BEC crossover, the critical field $H_c=H_c(\mu)=\gamma\mu$ for the first
order phase transition in the grand canonical ensemble splits into
a lower critical field $H_{c1}$ and an upper one $H_{c2}$, and the
SF-N mixed phase should appear in the interval $H_{c1}<H<H_{c2}$. This can be
understood by the fact that the chemical potentials for the normal and superfluid phases
do not equate in the BCS-BEC crossover, which is different from the classical weak coupling case where
the chemical potential is always set be equal to the Fermi energy $\epsilon_{\text F}$.

$H_{c1}$ and $H_{c2}$ can be determined by setting the chemical potential
$\mu$ to be its value in the superfluid and the normal phases,
respectively. We have $H_{c1}=\gamma\mu_{\text{SF}}(H_{c1})$ and
$H_{c2}=\gamma\mu_{\text N}(H_{c2})$, where $\mu_{\text{SF}}(H_{c1})=\xi\epsilon_{\text F}$. $\mu_{\text
N}(H_{c2})$ can be obtained by the number equation (\ref{numn}) of the normal
state at $H=H_{c2}$. We find
\begin{equation}
\mu_{\text N}(H_{c2})=\frac{\epsilon_{\text
F}}{\left[\xi^{-3/2}-\frac{2}{5}\gamma {\cal
G}^\prime(\gamma)\right]^{2/3}},
\end{equation}
where we have used the fact ${\cal G}(\gamma)=\xi^{-3/2}$. Therefore, we arrive at the following
model-independent expressions for the lower and upper critical
Zeeman fields
\begin{eqnarray}
H_{c1}&=&\gamma\xi\epsilon_{\text F},\nonumber\\
H_{c2}&=&\frac{\gamma\xi}{\left[1-\frac{2}{5}\xi^{3/2}\gamma {\cal
G}^\prime(\gamma)\right]^{2/3}}\epsilon_{\text
F}. \label{CC}
\end{eqnarray}
The appearance of the mixed phase requires
$H_{c1}<H_{c2}$, which gives rise to
\begin{equation}
{\cal
G}^\prime(\gamma)>0,\ \ \ \gamma {\cal
G}^\prime(\gamma)<\frac{5}{2}\xi^{-3/2}.
\end{equation}
Furthermore, since the superfluid
phase is unpolarized, the condition $H_{c1}<\Delta_0$ should be satisfied,
which leads to the fact $\gamma<\beta$. Therefore, the QMC result $\beta\simeq1.2$ \cite{gap01,gap02} gives
an upper bound for $\gamma$.

If we assume that the normal phase is fully polarized, the two critical fields can determined
with only one parameter $\xi$,
\begin{eqnarray}
H_{c1}&=&\left(\frac{2^{2/5}}{\xi^{3/5}}-1\right)\xi\epsilon_{\text F},\nonumber\\
H_{c2}&=&\left(\frac{2^{2/5}}{\xi^{3/5}}-1\right)\frac{2^{4/15}}{\xi^{3/5}}\epsilon_{\text
F}.
\end{eqnarray}
Substituting $\xi=0.42$, we obtain $H_{c1}=0.51\epsilon_{\text F}$ and $H_{c2}=2.47\epsilon_{\text F}$.
However, this is not the realistic case and is also not consistent with the QMC result for the pairing gap
$\Delta_0=0.4-0.5\epsilon_{\text F}$ \cite{gap01,gap02}.

%%%%%%%%%%%%%%%%%%%%%%%%%%%%%%%%%%%%%%%%%%%%%%%%%%%%%%%%%%%%%%%%%%%%%%%
\subsection {Properties of the Mixed Phase}
\label{s2-3}
%%%%%%%%%%%%%%%%%%%%%%%%%%%%%%%%%%%%%%%%%%%%%%%%%%%%%%%%%%%%%%%%%%%%%%%
A SF-N mixed phase should appear in the interval
$H_{c1}<H<H_{c2}$, which essentially corresponds to the one observed in QMC calculations at $0<P<P_c$ \cite{MC}.
From the phase equilibrium condition ${\cal
P}_{\text{SF}}(\mu,H)={\cal P}_{\text N}(\mu,H)$, in the mixed
phase the ratio $H/\mu$ keeps a constant $\gamma$ and the chemical
potential reads $\mu_{\text{M}}(H)=H/\gamma$. The properties of the
normal domain in the mixed phase depend on the value of $\gamma$.
The QMC calculations~\cite{MC} and experimental data support $\gamma<\delta_0$, i.e., the
normal domain is partially polarized. The volume fractions of the
superfluid and normal phases can be denoted by $y$ and $1-y$, respectively. $y(H)$ is determined by the equation
$n=y(H)n_{\text{SF}}(\mu_{\text{M}},H)+[1-y(H)]n_{\text
N}(\mu_{\text{M}},H)$. Using the EOS for the phases SF and N, we
find
\begin{equation}
y(H)=\frac{5\xi^{-3/2}}{2\gamma
{\cal G}^\prime(\gamma)}\left[\left(\frac{H}{H_{c1}}\right)^{-3/2}-\left(\frac{H_{c2}}{H_{c1}}\right)^{-3/2}\right].
\end{equation}
The energy density of the mixed phase, ${\cal
E}_{\text{M}}(H)=\mu_{\text{M}}n-{\cal
P}_{\text{M}}(\mu_{\text{M}},H)$, can be evaluated as
\begin{equation}
{\cal E}_{\text{M}}(H)
=\frac{5}{3}\frac{H}{H_{c1}}\left[1-\frac{2}{5}\left(\frac{H}{H_{c1}}\right)^{3/2}\right]\xi{\cal
E}_0.
\end{equation}
Since the normal domain possesses imbalanced spin populations, there is a nonzero global
polarization $P$ in the mixed phase, i.e., the system becomes
spin-polarized when $H>H_{c1}$. From the definition of $P$,
we find
\begin{equation}
P(H)=\frac{1}{\gamma}\left[\left(\frac{H}{H_{c1}}\right)^{3/2}-1\right],\ \ \ H_{c1}<H<H_{c2}.
\end{equation}
The mixed phase continuously links the superfluid and normal phases.
One can easily show that $\mu_{\text{M}}=\mu_{\text{SF}}$ at $H=H_{c1}$ and
$\mu_{\text{M}}=\mu_{\text{N}}$ at $H=H_{c2}$, which ensures
$0\leq y\leq 1$ with $y(H_{c1})=1$ and $y(H_{c2})=0$.

\begin{figure}[!htb]
\begin{center}
\includegraphics[width=8.5cm]{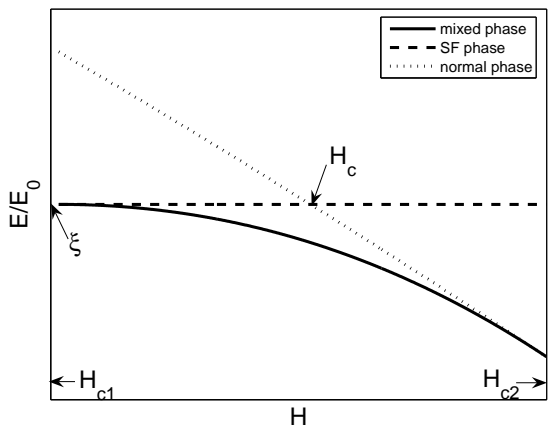}
\caption{A schematic plot of the energy densities (divided by the energy density of the ideal Fermi gas) for the superfluid, normal and mixed phases in
the region $H_{c1}<H<H_{c2}$. \label{fig2}}
\end{center}
\end{figure}

For the discussions above, the mixed phase is assumed to be the
ground state in the region $H_{c1}<H<H_{c2}$. As a complete study,
we have to prove that the mixed phase has the lowest energy in
this region. Even though the full information of the scaling function ${\cal
G}(x)$ is still not clear, this can be done if the function ${\cal G}(x)$
behaves sufficiently regularly.
\\
(1) The energy density of the mixed phase can be written as
${\cal E}_{\text{M}}(H)=\frac{5}{3}\xi{\cal E}_0f(H/H_{c1})$ with
$f(z)=z-\frac{2}{5}z^{5/2}$. From $f^\prime(z)=1-z^{3/2}$, ${\cal
E}_{\text{M}}(H)$ is a monotonously decreasing function of $H$ in
the region $H_{c1}<H<H_{c2}$. Combining ${\cal E}_{\text{M}}={\cal E}_{\text{SF}}$
at $H=H_{c1}$ and the fact that ${\cal
E}_{\text{SF}}$ is $H$-independent, there is always ${\cal
E}_{\text{M}}(H)<{\cal E}_{\text{SF}}(H)$ for $H_{c1}<H<H_{c2}$.
\\
(2) The condition ${\cal E}_{\text{M}}(H)<{\cal
E}_{\text{N}}(H)$ requires $g(\gamma)<g(\gamma^\prime)$ with
$g(t)=h/t-\frac{2}{5}{\cal G}(t)\left(h/t\right)^{5/2}$,
$h=H/\epsilon_{\text F}$ and $\gamma^\prime=H/\mu_{\text N}(H)$.
Even though we lack the full information of the scaling function ${\cal
G}(x)$, it is sufficient to show ${\cal
E}_{\text{M}}(H)<{\cal E}_{\text{N}}(H)$ at $H\lesssim H_{c2}$,
due to the continuity and the BCS instability ${\cal
E}_{\text{N}}(0)>{\cal E}_{\text{SF}}(0)$. From the first-order
derivative of $g(t)$ at $t=\gamma$,
$g^\prime(\gamma)=\gamma^{-2}h\left[(H/H_{c2})^{3/2}-1\right]$,
$g(t)$ is a decreasing function near $t=\gamma$. Therefore, at
$H\lesssim H_{c2}$ the condition $g(\gamma)<g(\gamma^\prime)$
requires $\gamma^\prime<\gamma$ or $\mu_{\text
N}>\mu_{\text{M}}$. From $\mu_{\text N}(0)>\mu_{\text{SF}}$ and
$\mu_{\text{M}}$ being an increasing function of $H$, the
relation $\mu_{\text{SF}}<\mu_{\text{M}}<\mu_{\text N}$ holds in
the region $H_{c1}<H<H_{c2}$. Therefore, the condition ${\cal E}_{\text{M}}(H)<{\cal
E}_{\text{N}}(H)$ is satisfied once the function ${\cal G}(x)$ behaves sufficiently regularly.
A schematic plot of the energy
densities for various phases is shown in Fig.\ref{fig2}.
%%%%%%%%%%%%%%%%%%%%%%%%%%%%%%%%%%%%%%%%%%%%%%%%%%%%%%%%%%%%%%%%%%%%%%%
%%%%%%%%%%%%%%%%%%%%%%%%%%%%%%%%%%%%%%%%%%%%%%%%%%%%%%%%%%%%%%%%%%%%%%%%

%%%%%%%%%%%%%%%%%%%%%%%%%%%%%%%%%%%%%%%%%%%%%%%%%%%%%%%%%%%%%%%%%%%%%%%
\section {Results and Comparison}
\label{s3}
%%%%%%%%%%%%%%%%%%%%%%%%%%%%%%%%%%%%%%%%%%%%%%%%%%%%%%%%%%%%%%%%%%%%%%%
Now we turn to determine the critical
Zeeman fields $H_{c1}$ and $H_{c2}$. Since the assumption of fully polarized normal phase is not
correct for $\gamma<\delta_0$ and $P_c<1$, we may have to know the full information of the
function ${\cal G}(x)$ in the partially polarized normal phase. Fortunately, in the following
we will find that $H_{c1}$ and $H_{c2}$ can be completely determined once the values of $\gamma$ and
$P_c$ (as well as $\xi$) are known.

From the exact expression (\ref{CC}), the lower critical field $H_{c1}$ can be determined once the values of
$\xi$ and $\gamma$ are known. On the other hand, to determine the
upper critical field $H_{c2}$, we need the value of ${\cal
G}^\prime(\gamma)$ which is not known so far. However, we find that we can determine the value of
${\cal G}^\prime(\gamma)$ from the critical population imbalance
$P_c=P(H_{c2})$ which can be determined in experiments and QMC calculations \cite{MC,exp}. To this end, we consider
a system with separately fixed $N_\uparrow$ and $N_\downarrow$ where the conversion between the two spin components
is forbidden. From
$P(H_{c1})=0$ and $P(H_{c2})=P_c$, the ground state is the
unpolarized superfluid state at $P=0$ and the SF-N mixed phase for
$0<P<P_c$. In the mixed phase, the effective ``magnetic field" is
given by $H(P)=\gamma\xi\epsilon_{\text F}(1+\gamma P)^{2/3}$, and
the critical population imbalance $P_c$ reads
\begin{equation}
P_c=\frac{\frac{2}{5}{\cal G}^\prime(\gamma)}{\xi^{-3/2}-\frac{2}{5}\gamma {\cal
G}^\prime(\gamma)}. \label{PC}
\end{equation}
Note that this expression is consistent with the jump in Fig. \ref{fig1} which can be
calculated from Eq. (4). Therefore, the upper critical field $H_{c2}$ can be determined once $P_c$ is known.
Further, the procedure can be simplified if we combine the expressions for the critical fields and the population imbalance,
i.e. Eqs. (\ref{CC}) and (\ref{PC}). In this way, we obtain a very simple relation among the CC limits for different cases
\begin{equation}
\frac{H_{c2}}{H_{c1}}=\left(1+\gamma P_c\right)^{2/3}. \label{CC2}
\end{equation}

For the case of fixed atom numbers, the energy density
defined as ${\cal
E}=\mu_{\uparrow}n_\uparrow+\mu_{\downarrow}n_\downarrow-{\cal P}$
satisfies the relation ${\cal E}=3{\cal P}/2$ in all phases, since
$H$ is now no longer treated as an external field. To show the consistency between the theoretical formula of $P_c$ obtained
above and the QMC calculations
and cold atom experiments we derive
the energy density ${\cal E}_{\text{M}}$ as a function of the
ratio $n_{\downarrow}/n_{\uparrow}$~\cite{MC}
\begin{equation}
{\cal E}_{\text{M}}(n_{\uparrow},n_{\downarrow})=
\frac{3}{5}n_{\uparrow}\frac{(6\pi^2n_{\uparrow})^{2/3}}{2M}I\left(\frac{n_{\downarrow}}{n_{\uparrow}}\right),
\end{equation}
where the function
$I(z)$ can be shown to be
$I(z)=2^{-2/3}\xi\left[(1+\gamma)+(1-\gamma)z\right]^{5/3}$.
The function
$I(z)$ is consistent with the formula used to obtain $P_c$ in the QMC
calculations \cite{MC}.

%%%%%%%%%%%%%%%%%%%%%%%%%%%%%%%%%%%%%%%%%%%%%%%%%%%%%%%%%%%%%%%%%%%%%
\begin{table}[b!]
\begin{center}
\begin{tabular}{|c| c c c | c c c |}
\hline
&&&&&&\\[-3mm]
 & $\gamma$  & $P_c$ & $\xi$ & ${\cal G}^\prime(\gamma)$ & $H_{c1}$ [$\epsilon_{\text F}$] & $H_{c2}$ [$\epsilon_{\text F}$]
\\[1mm]
\hline
&&&&&&\\[-3mm]
QMC \cite{MC}& 0.967 & 0.389 & 0.42 &2.596 & 0.406 & 0.503\\
MIT \cite{exp}& 0.95 & 0.36 & 0.42 &2.464 &0.399 &0.485 \\
LKB-Lhomond \cite{exp2}& 0.878 & 0.324 & 0.42 &2.317 &0.369 &0.436\\
LKB-Lhomond \cite{exp3}& 0.897 & 0.359 & 0.41 &2.586 &0.368 &0.443 \\
\hline
\end{tabular}
\end{center}
\caption{\small The data of $\gamma$, $P_c$ and $\xi$ from the QMC study \cite{MC} and experiments \cite{exp,exp2,exp3}
and the values of the critical Zeeman fields $H_{c1}$ and $H_{c2}$ determined from the formulae (\ref{CC}) and (\ref{CC2}).
The value of ${\cal G}^\prime(\gamma)$ determined from Eq. (\ref{PC}) is also shown. The value of $P_c$ in \cite{exp3} is not given,
and we use our formula to extract it from the experimental data,  see Fig. \ref{fig3}.} \label{Zeeman}
\end{table}
%%%%%%%%%%%%%%%%%%%%%%%%%%%%%%%%%%%%%%%%%%%%%%%%%%%%%%%%%%%%%%%%%%%%%

\begin{figure}[!htb]
\begin{center}
\includegraphics[width=9cm]{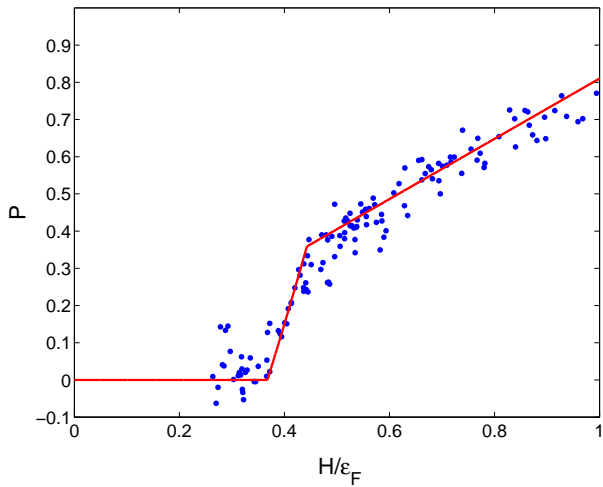}
\caption{(color online) The spin polarization $P=(N_\uparrow-N_\downarrow)/(N_\uparrow+N_\downarrow)$ as a function of
the external Zeeman field $H$ (divided by $\epsilon_{\text F}$) for a homogeneous Fermi gas at unitary. The blue dots are data
extracted from the LKB-Lhomond experiments~\cite{exp3,note}. The red solid line is the theoretical prediction. We use the data $\gamma=0.897$, $\xi=0.41$ from
\cite{exp3} and $H_{c1}=0.368\epsilon_{\text F}$. For the normal phase $H>H_{c2}$, we use the linear fit $P=\frac{3}{2}\tilde{\chi} H/\epsilon_{\text F}$
with $\tilde{\chi}=0.54$ suggested in \cite{exp3}. In between, the formula $P=\gamma^{-1}[(H/H_{c1})^{3/2}-1]$ is adopted. The value of $P_c$ is self-consistently
determined as $P_c=0.359$.  \label{fig3}}
\end{center}
\end{figure}

We can now determine the critical Zeeman fields $H_{c1}$ and $H_{c2}$
from the known values of $\xi,\gamma$ and $P_c$ extracted from QMC calculations and experimental measurements.
The results are listed in Table~\ref{Zeeman}. In Fig.~\ref{fig3} we show the data
of the polarization $P$ as a function of the Zeeman splitting $H/\epsilon_{\text F}$ extracted from the LKB-Lhomond experiments~\cite{exp3}.
The critical Zeeman fields $H_{c1}$ and $H_{c2}$ are around $0.4\epsilon_{\text F}$ and consistent with our calculations in Table \ref{Zeeman}
from the LKB-Lhomond data. The deviation between the results from LKB-Lhomond data and those from QMC and MIT data comes mainly from the
difference in the values of $\gamma$. In the early studies~\cite{MC,exp}, the value of $\gamma$ was reported to be around $0.96$. However, recent data from
LKB-Lhomond experiments show that this value becomes smaller, around $0.89$.

%%%%%%%%%%%%%%%%%%%%%%%%%%%%%%%%%%%%%%%%%%%%%%%%%%%%%%%%%%%%%%%%%%%%%%%
%%%%%%%%%%%%%%%%%%%%%%%%%%%%%%%%%%%%%%%%%%%%%%%%%%%%%%%%%%%%%%%%%%%%%%%%

%%%%%%%%%%%%%%%%%%%%%%%%%%%%%%%%%%%%%%%%%%%%%%%%%%%%%%%%%%%%%%%%%%%%%
\begin{table}[b!]
\begin{center}
\begin{tabular}{|c| c c | c c |}
\hline
&&&&\\[-3mm]
& $P_c$ & $H_{c2}$ [$\epsilon_{\text F}$] & $\chi/\chi_0$ & $H_{0}$ [$\epsilon_{\text F}$]
\\[1mm]
\hline
&&&&\\[-3mm]
QMC \cite{MC}& 0.389 & 0.503 & 0.52 &1.29 \\
MIT \cite{exp}& 0.36 & 0.485 & 0.49 &1.35 \\
LKB-Lhomond \cite{exp2}& 0.324 & 0.436 & 0.50 &1.35 \\
LKB-Lhomond \cite{exp3}& 0.359 & 0.443 & 0.54 &1.23 \\
\hline
\end{tabular}
\end{center}
\caption{\small The spin susceptibility $\chi$ (divided by $\chi_0$) and the Zeeman splitting $H_0$ at which the Fermi gas becomes fully polarized
determined from different QMC and experimental data. } \label{SUS}
\end{table}
%%%%%%%%%%%%%%%%%%%%%%%%%%%%%%%%%%%%%%%%%%%%%%%%%%%%%%%%%%%%%%%%%%%%%

It was shown in \cite{exp2,exp3} that the normal phase of the unitary Fermi gas is well described by Landau's Fermi liquid theory.
Here we can extract the spin susceptibility $\chi$ of the normal phase from the determined upper critical field $H_{c2}$. The spin susceptibility
$\chi$ is defined by the linear relation between the spin polarization $P$ and the Zeeman splitting $H$, i.e., $P=(3\chi H)/(2\chi_0\epsilon_{\text F})$ where
$\chi_0=3n/2\epsilon_{\text F}$ is the spin susceptibility of an ideal Fermi gas. Therefore, the spin susceptibility $\chi$ can be determined by
\begin{equation}
\frac{\chi}{\chi_0}=\frac{P_c}{\frac{3}{2}\frac{H_{c2}}{\epsilon_{\text F}}}.
\end{equation}
We can also determine the saturation splitting $H_0$ at which the Fermi gas becomes fully polarized by setting $P=1$. We find $H_0/\epsilon_{\text F}=2\chi_0/3\chi$.
The numerical results for $\chi$ and $H_0$ from different QMC and experimental data are summarized in Table~\ref{SUS}. We find good agreement
among the results. The spin susceptibility of this unitary Fermi liquid is about half of that of the ideal Fermi gas, and the saturation splitting $H_0$ is around
$1.3\epsilon_{\text F}$.

Finally, we compare our results with
those obtained from the mean-field~\cite{sheehy} and beyond-mean-field~\cite{Vei}
theories.
Our formulae (\ref{CC}) and (\ref{PC}) for the CC limits are model independent. In the
mean-field theory, the universal constant $\xi$ reads $\xi_{\text{MF}}=0.5906$ and the scaling function ${\cal G}(x)$ takes the non-interacting form:
${\cal G}_{\text{MF}}(x)=\frac{1}{2}\left[(1+x)^{5/2}\Theta(1+x)+(1-x)^{5/2}\Theta(1-x)\right]$.
Numerical solution of equation (\ref{gr}) leads to
$\gamma_{\text{MF}}=0.8071$ and therefore ${\cal
G}_{\text{MF}}^\prime(\gamma_{\text{MF}})=2.9307$. Substituting these values into (\ref{CC}) and (\ref{PC}) we obtain
\begin{equation}
H_{c1}^{\text{MF}}=0.477\epsilon_{\text F},\ \ H_{c2}^{\text{MF}}=0.693\epsilon_{\text F},\ \ P_c^{\text{MF}}=0.933,
\end{equation}
which agree well with the numerical values
obtained in~\cite{sheehy}. One finds that the mean field results of the critical Zeeman fields
$H_{c1}$ and $H_{c2}$ deviate significantly from our results listed in Table \ref{Zeeman}. The beyond-mean-theories may
properly include the pairing fluctuations in the superfluid phase and the interactions in the normal phase
and give reasonable values of the critical Zeeman fields. Such a calculation
for the upper critical field $H_{c2}$ within the large-N expansion method has been performed in \cite{Vei}. To the leading order of $1/N$,
the result reads
\begin{equation}
\frac{H_{c2}}{\epsilon_{\text F}}=0.693+\frac{0.087}{N}+O\left(\frac{1}{N^2}\right).
\end{equation}
It is clear that the leading order result from the large-N expansion
method is not consistent with our result, and the higher order calculations are needed.

%%%%%%%%%%%%%%%%%%%%%%%%%%%%%%%%%%%%%%%%%%%%%%%%%%%%%%%%%%%%%%%%%%%%%
\begin{table}[b!]
\begin{center}
\begin{tabular}{|c| c | c | c |}
\hline
&&&\\[-3mm]
 neutron density  & $n=10^{-3}n_0$  & $n=10^{-2}n_0$ & $n=10^{-1}n_0$
\\[1mm]
\hline
&&&\\[-3mm]
$(k_{\text F}a_{nn})^{-1}$ & -0.32 & -0.15 & -0.07
\\[1mm]
\hline
&&&\\[-3mm]
$B_{c1}$ [Gauss] & $3.6\times10^{16}$ & $1.7\times10^{17}$ & $7.7\times10^{17}$
\\[1mm]
\hline
&&&\\[-3mm]
$B_{c2}$ [Gauss]& $4.3\times10^{16}$ & $2.0\times10^{17}$ & $9.3\times10^{17}$  \\
\hline
\end{tabular}
\end{center}
\caption{\small The lower and upper critical magnetic fields, $B_{c1}$ and $B_{c2}$, for dilute neutron matter at
different densities, $n/n_0=10^{-3},10^{-2},10^{-1}$. The values of $(k_{\text F}a_{nn})^{-1}$ are also shown,
from which we find that the effective couplings at these densities are really close to the unitary limit. In the calculations,
we adopt the data $\xi=0.41$, $\gamma=0.897$ and $P_c\simeq0.36$ from \cite{exp3}.} \label{neutron}
\end{table}
%%%%%%%%%%%%%%%%%%%%%%%%%%%%%%%%%%%%%%%%%%%%%%%%%%%%%%%%%%%%%%%%%%%%%

%%%%%%%%%%%%%%%%%%%%%%%%%%%%%%%%%%%%%%%%%%%%%%%%%%%%%%%%%%%%%%%%%%%%%%%
\section {Indication to Dilute Neutron Matter}
\label{s4}
%%%%%%%%%%%%%%%%%%%%%%%%%%%%%%%%%%%%%%%%%%%%%%%%%%%%%%%%%%%%%%%%%%%%%%%
For neutron
matter the effective range of the nuclear force, $r_0\simeq2.7$fm, is much smaller than the s-wave neutron-neutron scattering
length, $a_{nn}\simeq-18.5$fm. Therefore, for dilute neutron matter which may exist in the crust of neutron stars,
$k_{\text F}r_0$ can be relatively small but $k_{\text F}a_{nn}$ remains large. So the properties of the dilute neutron matter
are close to the unitary Fermi gas discussed in the paper.

As a naive application, we can estimate the
critical magnetic fields for the dilute neutron matter at which the superfluid state is destroyed and the matter becomes spin-polarized.
The Zeeman splitting $E_Z=2H$
in this case is $H=\mu_n B$ where $\mu_n$ is the magnetic moment of the neutrons and $B$ the magnetic field. After some simple
algebras, we obtain the following formulae for the lower and upper magnetic fields
\begin{eqnarray}
B_{c1}&=&\gamma\xi\left(\frac{n}{n_0}\right)^{2/3}B_0,\nonumber\\
B_{c2}&=&\gamma\xi(1+\gamma P_c)^{2/3}\left(\frac{n}{n_0}\right)^{2/3}B_0,
\end{eqnarray}
with $n_0=0.16$fm$^{-3}$ being the nuclear saturation density and $B_0=(3\pi^2n_0)^{2/3}/(2M_n\mu_n)\simeq10^{19}$G. In Table \ref{neutron},
we calculate the critical magnetic fields $B_{c1}$ and $B_{c2}$ for some typical densities of dilute neutron matter from $n/n_0=10^{-3}$ to
$n/n_0=10^{-1}$. We find that the critical magnetic fields are roughly in the range $10^{16}-10^{18}$Gauss.
Therefore, the problem of imbalanced pairing and spin-polarization in dilute neutron matter is in
principle relevant to compact objects known as magnetars \cite{magnetar}, which have surface magnetic fields of $10^{14}-10^{15}$G \cite{mag01}.
In fact, according to the scalar virial theorem which is based on Newtonian gravity, the magnetic
field strength is allowed by values up to $10^{18}$G in the interior of a magnetar \cite{mag02,mag03,mag04}.

%%%%%%%%%%%%%%%%%%%%%%%%%%%%%%%%%%%%%%%%%%%%%%%%%%%%%%%%%%%%%%%%%%%%%%%
\section {Extension to Nonzero Temperature}
\label{s5}
%%%%%%%%%%%%%%%%%%%%%%%%%%%%%%%%%%%%%%%%%%%%%%%%%%%%%%%%%%%%%%%%%%%%%%%
The above model-independent approach can be generalized to finite
temperature $T$, where both the normal and superfluid phases are
spin-polarized due to the thermal excitations quasi-particles. From the universality, the
EOS for the normal and superfluid phases read \cite{ho}
\begin{equation}
{\cal P}_{{\text N},\text{SF}}(T,\mu,H)={\cal P}_0(\mu){\cal
G}_{{\text N},\text{SF}}\left(\frac{H}{\mu},\frac{T}{\mu}\right),
\end{equation}
where we have set the Boltzmann constant $k_{\text B}=1$. The scaling functions for the
normal and superfluid phases should be different.

In the grand canonical ensemble, one expects that the phase
transition along the $T/\mu$ axis is of second order at small
$H/\mu$ and first order at large $H/\mu$. The first order phase
transition is determined by the equation ${\cal G}_{\text
N}\left(H/\mu,T/\mu\right)={\cal
G}_{\text{SF}}\left(H/\mu,T/\mu\right)$, or explicitly
$H/\mu={\cal W}(T/\mu)$ with known ${\cal W}(0)=\gamma$. The first
order phase transition should end at a so-called tricritical point
$(H/\mu,T/\mu)=(a,b)$. At the mean-field level, it is predicted to be
$(a,b)=(0.70,0.38)$~\cite{parish}.

At fixed total particle number, $\mu$ is not a free variable, and
the tricritical point is characterized by
$(T_{\text{TCP}},H_{\text{TCP}})$. Due to the continuity with the
zero temperature case, for $T<T_{\text{TCP}}$, there exist two
critical fields $H_{c1}(T)=\mu_1{\cal W}(T/\mu_1)$ and
$H_{c2}(T)=\mu_2{\cal W}(T/\mu_2)$, where $\mu_1$ and $\mu_2$ are
the chemical potentials corresponding to the superfluid phase at $H=H_{c1}$ and the normal phase at $H=H_{c2}$, respectively.
The region $H_{c1}<H<H_{c2}$ for the mixed phase should decrease with
increasing $T$, and finally disappear at the tricritical point
where $H_{\text{TCP}}=aT_{\text{TCP}}/b$ and
$\mu_1=\mu_2=T_{\text{TCP}}/b$.

When $N_\uparrow$ and $N_{\downarrow}$ are fixed, for
$T<T_{\text{TCP}}$, the SF-N mixed phase should be the ground
state in the region $P_1<P<P_2$ with $P_1=P(H_{c1})$ and
$P_2=P(H_{c2})$. At $T\neq 0$, $P_1$ should be nonzero and
increase with the temperature. At the
tricritical point, $P_1=P_2=P_{\text{TCP}}$ is satisfied. Once the scaling function ${\cal G}$ and the
tricritical point $(a,b)$ in the grand canonical ensemble are known, $P_{\text{TCP}}$ and
$T_{\text{TCP}}$ can be calculated from the following
model-independent formulae
\begin{eqnarray}
P_{\text{TCP}}&=&\frac{{\cal G}_x^\prime(a,b)}{\frac{5}{2}{\cal
G}(a,b)-a{\cal G}_x^\prime(a,b)-b{\cal
G}_y^\prime(a,b)},\nonumber\\
\frac{T_{\text{TCP}}}{\epsilon_{\text
F}}&=&b\left[\frac{5P_{\text{TCP}}}{2{\cal
G}_x^\prime(a,b)}\right]^{2/3}
\end{eqnarray}
with the definition ${\cal G}_x^\prime(x,y)=\partial{\cal
G}(x,y)/\partial x$ and ${\cal G}_y^\prime(x,y)=\partial{\cal
G}(x,y)/\partial y$, where ${\cal G}$ can be the scaling function of
either the superfluid or the normal phase.

%%%%%%%%%%%%%%%%%%%%%%%%%%%%%%%%%%%%%%%%%%%%%%%%%%%%%%%%%%%%%%%%%%%%%%%
\section {Summary}
\label{s6}
%%%%%%%%%%%%%%%%%%%%%%%%%%%%%%%%%%%%%%%%%%%%%%%%%%%%%%%%%%%%%%%%%%%%%%%

In summary, we have determined the lower and upper critical Zeeman fields for a homogeneous
Fermi superfluid at infinite scattering length. Using the recent experimental data from LKB-Lhomond,
we found $H_{c1}\simeq0.37\epsilon_{\text F}$ and $H_{c2}\simeq0.44\epsilon_{\text F}$.
The value of the lower critical field also gives a lower bound for the excitation gap $\Delta_0$ for the balanced case.
The results are highly related to the
properties of dilute neutron matter in presence of a strong magnetic field which may exists in compact objects.
Theoretically, it is interesting that
we can prove the existence of the two critical fields~($H_{c1}$ and $H_{c2}$) based only on the universal equations of state.
We also presented a very simple proof for the fact that the mixed phase has the lowest energy, in contrast to the
proof for the weak coupling case \cite{BCR}. The value of the universal constant $\xi$ and the properties of the polarized normal phase are
very important in calculating the
CC limits. Future theoretical studies may focus on the calculation
of the scaling function ${\cal G}(x)$ from beyond-mean-field
theories~\cite{Vei,Hu,Son}. Once the universal constant $\xi$ and the
function ${\cal G}(x)$ are known, one can directly obtain the
critical polarization $P_c$ from our model-independent formula
(\ref{PC}) and check the consistency between theories and experiments
or QMC calculations.

{\bf Acknowledgments:}\ We thank Sylvain  Nascimb\`{e}ne for providing us with the experimental data.
L. He thanks the support from the Alexander von Humboldt Foundation, and P. Zhuang is supported by the NSFC Grants
10735040, 10975084 and 11079024.


\begin{thebibliography}{99}
\bibitem{BCS}
J. R. Schrieffer, \emph{Theory of
Superconductivity}, Adddison-Wesley, 1964.
\bibitem{BCSBEC}
A. J. Leggett, in \emph{Modern trends in the theory of condensed
matter}, Springer-Verlag, Berlin, 1980.
\bibitem{BCSBEC02}
D. M. Eagles, Phys. Rev. {\bf 186}, 456(1969).
\bibitem{BCSBEC03}
P. Nozi\`{e}res and S. Schmitt-Rink, J. Low Temp. Phys. {\bf 59}, 195(1985).
\bibitem{BCSBEC04}
C. A. R. S\'{a} de Melo, M. Randeria, and J. R. Engelbrecht, Phys. Rev. Lett. {\bf 71}, 3202(1993).
\bibitem{BCSBECexp}
M. Greiner, C. A. Regal, and D. S. Jin, Nature {\bf 426}, 537(2003).
\bibitem{BCSBECexp02}
S. Jochim, M. Bartenstein, A. Altmeyer, G. Hendl, S. Riedl, C. Chin, J. H. Denschlag, and R. Grimm, Science {\bf 302}, 2101(2003).
\bibitem{BCSBECexp03}
M. W. Zwierlein, J. R. Abo-Shaeer, A. Schirotzek, C. H. Schunck, and W. Ketterle, Nature {\bf 435}, 1047(2003).
\bibitem{ho}
T. -L. Ho, Phys. Rev. Lett. {\bf 92}, 090402(2004).
\bibitem{DNM}
J. Carlson, J. Morales, Jr., V. R. Pandharipande, and D. G. Ravenhall, Phys. Rev. {\bf C68}, 025802(2003).
\bibitem{DNM02}
S. Y. Chang, J. Morales, Jr., V. R. Pandharipande, D. G. Ravenhall, J. Carlson, S. C. Pieper, R. B. Wiringa, and K. E. Schmidt,
Nucl. Phys. {\bf A746}, 215(2004).
\bibitem{DNM03}
A. Gezerlis and J. Carlson, Phys. Rev. {\bf C77}, 032801(2008); Phys. Rev. {\bf C81}, 025803(2010).
\bibitem{CC}
B. S. Chandrasekhar, Appl. Phys. Lett. {\bf 1}, 7(1962).
\bibitem{CC2}
A. M. Clogston, Phys. Rev. Lett. {\bf 9}, 266(1962).
\bibitem{Sarma}
G. Sarma, J. Phys. Chem. Solid {\bf 24},1029(1963).
\bibitem{FFLO}
P. Fulde and R. A. Ferrell, Phys. Rev {\bf 135}, A550(1964).
\bibitem{FFLO2}
A. I. Larkin and Yu. N. Ovchinnikov, Sov. Phys. JETP {\bf 20},
762(1965).
\bibitem{imbalanceexp}
M. W. Zwierlein, A. Schirotzek, C. H. Schunck, and W. Ketterle, Science {\bf 311}, 492(2006).
\bibitem{imbalanceexp2}
G. B. Partridge, W. Li, R. I. Kamar, Y. -an Liao, and R. G. Hulet, Science {\bf 311}, 503(2006).
\bibitem{BCR}
P. F. Bedaque, H. Caldas, and G. Rupak, Phys. Rev. Lett. {\bf 91}, 247002(2003).
\bibitem{Cohen}
T. D. Cohen, Phys. Rev. Lett. {\bf 95}, 120403(2005).
\bibitem{Carlson}
J. Carlson and S. Reddy, Phys. Rev. Lett. {\bf 95}, 060401(2005).
\bibitem{BP}
W. V. Liu and F. Wilczek, Phys. Rev. Lett. {\bf 90}, 047002(2003);
E. Gubankova, W. V. Liu, and F. Wilczek, Phys. Rev. Lett. {\bf 91}, 032001(2003);
M. M. Forbes, E. Gubankova, W. V. Liu, and F. Wilczek, Phys. Rev. Lett. {\bf 94}, 017001(2005).
\bibitem{he1}
L. He, M. Jin, and P. Zhuang, Phys. Rev.{\bf B73}, 214527(2006); Phys. Rev. {\bf B74}, 024516(2006);
Phys. Rev. {\bf B74}, 214516(2006).
\bibitem{nuclear}
A. Sedrakian and U. Lombardo, Phys. Rev. Lett. {\bf 84}, 602(2000); A. Sedrakian, Phys. Rev. {\bf C63}, 025801(2001);
H. M{\"u}ther and A. Sedrakian, Phys. Rev. {\bf C67}, 015802(2003); M. Jin, L. He, and P. Zhuang, Int. J. Mod. Phys. {\bf E16},
2363(2007).
\bibitem{quark}
M. Huang, P. Zhuang, and W. Chao, Phys. Rev. {\bf D67}, 065015(2003);
I. Shovkovy and M. Huang, Phys. Lett. {\bf B564}, 205(2003);
M. Alford, C. Kouvaris, and K. Rajagopal, Phys. Rev. Lett. {\bf 92}, 222001(2004).
\bibitem{qmcn}
A. Gezerlis, arXiv:1012.4464.
\bibitem{MC}
C. Lobo, A. Recati, S. Giorgini, and S. Stringari, Phys. Rev. Lett. {\bf 97}, 200403(2006).
\bibitem{exp}
Y. Shin, C. H. Schunck, A. Schirotzek, and W. Ketterle, Nature {\bf 451}, 689(2008).
\bibitem{exp2}
S. Nascimb\`{e}ne, N. Navon, K. Jiang, F. Chevy, and C. Salomon, Nature {\bf 463}, 1057(2010).
\bibitem{exp3}
S. Nascimb\`{e}ne, N. Navon, S. Pilati, F. Chevy, S. Giorgini, A. Georges, and C. Salomon, arXiv:1012.4664.
\bibitem{note}
The data shown in Fig. \ref{fig3} can not be obtained by direct experimental measurements. In recent experiments \cite{exp2,exp3},
people probed the local pressure of the trapped gas using \emph{in situ} images, and then obtained the pressure ${\cal P}(\mu,H)$ for
homogeneous systems, according to the local density approximation. The spin polarization $P$ and the Fermi energy for a homogeneous
system can be obtained from the same universal thermodynamic relations used in this paper. For details, see \cite{exp2,exp3}.
\bibitem{sheehy}
D. E. Sheehy and L. Radzihovsky, Phys. Rev. Lett. {\bf 96},
060401(2006).
\bibitem{sheehy2}
D. E. Sheehy and L. Radzihovsky, Ann. Phys. (N. Y.){\bf 322}, 1790(2007).
\bibitem{he2}
L. He and P. Zhuang, Phys. Rev. {\bf A78}, 033613(2008).
\bibitem{forbes}
A. Bulgac and M. M. Forbes, Phys. Rev. {\bf A75}, 031605(2007).
\bibitem{chevy}
F. Chevy, Phys. Rev. Lett. {\bf 96}, 130401(2006); Phys. Rev. {\bf
A74}, 063628(2006).
\bibitem{son}
D. T. Son and M. A. Stephanov, Phys. Rev. {\bf A74}, 013614(2006).
\bibitem{fulmu}
R. Combescot, A. Recati, C. Lobo, and F. Chevy, Phys. Rev. Lett. {\bf 98}, 180402(2007).
\bibitem{gap01}
J. Carlson, S. -Y. Chang, V. R. Pandharipande, and K. E. Schmidt, Phys. Rev. Lett. {\bf 91}, 050401(2003);
S. -Y. Chang, V. R. Pandharipande, J. Carlson, and K. E. Schmidt, Phys. Rev. {\bf A70}, 043602(2004).
\bibitem{gap02}
J. Carlson and S. Reddy, Phys. Rev. Lett. {\bf 100}, 150403(2008).
\bibitem{parish}
M. M. Parish, F. M. Marchetti, A. Lamacraft, and B. D. Simons , Nat. Phys. {\bf 3}, 124(2007).
\bibitem{Vei}
M. Y. Veillette, D. E. Sheehy, and L. Radzihovsky, Phys. Rev. {\bf A75}, 043614(2007).
\bibitem{magnetar}
C. Kouveliotou, R. C. Duncan, and C. Thompson, Scientific American (February 2003), Page 35-41.
\bibitem{mag01}
C. Thompson and R. C. Duncan, Astrophys. J. {\bf 408}, 194(1993).
\bibitem{mag02}
D. Lai and S. L. Shapiro, Astrophys. J. {\bf 383}, 745(1991).
\bibitem{mag03}
S. Chakrabarty, D. Bandyopadhyay, and S. Pal, Phys. Rev. Lett. {\bf 78}, 2898(1997).
\bibitem{mag04}
A. Broderick, M. Prakash, and J. M. Lattimer, Astrophys. J. {\bf 537}, 351(2000).
\bibitem{Hu}
H. Hu, X. -J. Liu, and P. D. Drumond, Nat. Phys. {\bf 3}, 469(2007).
\bibitem{Son}
Y. Nishida and D. T. Son, Phys. Rev. Lett. {\bf 97}, 050403(2006).
\end{thebibliography}
\end{document}